# CALIBRATION OF THE EXTENDED QUADRAHEDRAL NaI(Tl) SPECTROMETER BY COSMIC MUONS

V.A. Baskov*, B. B. Govorkov†, V.V. Polyansky

*P.N. Lebedev Physical Institute (LPI),
Moscow, 119991 Russia, 53 Leninsky Prospekt*
*e-mail: baskov@x4u.lebedev.ru*

The results of the extended quadrahedral NaI(Tl) spectrometer`s calibration by cosmic muons are presented. The signal amplitude and energy resolution of the spectrometer have complex dependencies on the voltage at the photomultiplier and the entry point of the particle into the spectrometer in the transverse direction. When the voltage across the photomultiplier divider photomultiplier is U = 1150 V the homogeneous signals area is observed at about 90% of the spectrometer length, the energy resolution is about 7%.

**Keywords:** *calibration, NaI(Tl) spectrometer, energy resolution, cosmic muons.*

The process of neutral pions` photoproduction near the threshold on neutrons has not been studied experimentally ($\gamma + n \rightarrow \pi^0 + n$). The study of this process is supposed to be performed on a special installation using an ejected electron beam and a photon tagging system on the accelerator S-25R "Pakhra" at the Lebedev Physical Institute of the Russian Academy of Sciences (LPI) [1, 2].

For purposes of this task, it is assumed to be study the behavior of the total cross-section neutral pions` photoproduction on a neutron based on the installation containing extended quadrahedral NaI(Tl) spectrometers (Fig. 1) [3]. Since γ-quanta from the decay of $\pi^0$ mesons develop electromagnetic showers in the transverse direction in the proposed configuration of the spectrometer arrangement, the research of the spectrometers characteristics in this direction was carried out in more detail.

Each of the four spectrometers of the installation has a size of $12 \times 12 \times 45$ cm$^3$ and consists of three conglutinated together blocks of the NaI(Tl) scintillator with a size of $12 \times 12 \times 15$ cm$^3$. The radiation length in the transverse and longitudinal direction is $4.8 X_0$ and $18 X_0$, respectively ($X_0 = 2.5$ cm - the radiation length of NaI(Tl) (p = 3.67 g/cm$^3$)). The spectrometer is viewed from the end with a single photomultiplier PMT-52 with a standard voltage divider.

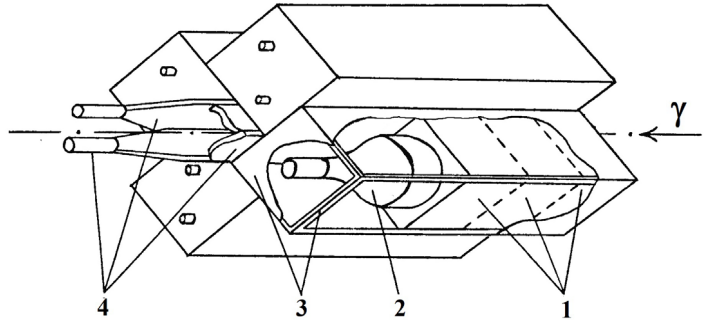

**Fig. 1.** The diagram of the set-up for studying the total photoproduction cross-section of neutral pions near the threshold on neutrons: 1-blocks of extended quadrahedral NaI(Tl) spectrometers; 2 – PMT-52; 3 – frame; 4 – scintillation counters of anticoincidence.

Preliminary calibration of the spectrometers was carried out on cosmic muons that create a minimum ionization of ~2 MeV·cm$^2$/g. The average energy left by cosmic muons in the transverse and longitudinal directions in the spectrometers was ~88 MeV and ~330 MeV, respectively.

The calibration scheme of one of the NaI(Tl) spectrometers is shown in Fig. 2. The spectrometer was positioned between two 4×4×0.5 cm$^3$ trigger counters $S_1$ and $S_2$, located at a distance of 15 cm from each other. Before calibration using the $^{60}$Co radiation source, the effective operating voltages on $S_1$ and $S_2$ were determined, and time delays were set between $S_1$ and $S_2$, as well as between $S_1$ and $S_2$ and the NaI(Tl) spectrometer.

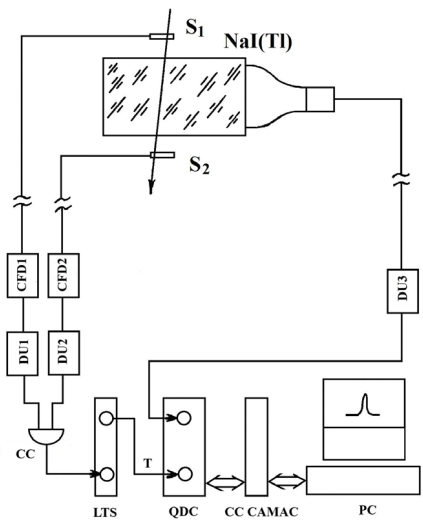

**Fig. 2.** The block diagram of calibration of the extended quadrahedral NaI(Tl) spectrometer: $S_1$ and $S_2$ – trigger scintillation counters; NaI(Tl) – extended quadrahedral spectrometer; CFD1 and CFD2 – constant-fraction discriminator; DU1 – DU3 – delay unit; CC – coincidence circuit; LTS – long time signal shaper; T – trigger (T = $S_1 \cdot S_2$); QDC – charge-to-digital converter; CC CAMAC – crate-controller CAMAC; PC - personal computer.

Signals from the passage of the cosmic muon through $S_1$ and $S_2$ were fed to the CFD1 and CFD2 – constant-fraction discriminators (the threshold for the formation of both discriminators was $U_{threshold}$ = 30 mV, the signals duration of the $T_{S1} = T_{S2}$ = 20 ns), then through the delay units DU1 and DU3 to the CC coincidence scheme. The signal from the CC coincidence scheme with a duration of $T_{CC}$ = 100 ns was not fed to the long-term signal forming unit LTS, which generated a signal T = 1000 ns. The signal T was a trigger signal (T = $S_1 \cdot S_2$) "Start" for actuation the unit 4 of the input charge-to-digital converter (QDC), which was used to "record" the signal from the NaI(Tl) spectrometer to the computer memory via the CAMAC system's crate-controller (CC CAMAC).

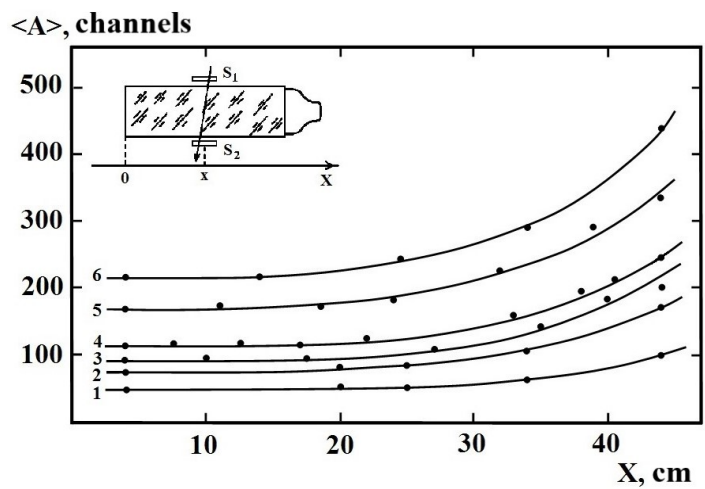

**Fig. 3.** Dependence of average amplitude of the signal of the extended quadrahedral NaI(Tl) spectrometer from the voltage on the voltage divider PMT-52: 1-U = 1100 V; 2-U = 1180 V; 3-U = 1217 V; 4-U = 1260 V; 5-U = 1305 V; 6-U = 1355 V.

The dependences of the average amplitude of the signal from the NaI(Tl) spectrometer on its length and voltage at the PMT are presented in Fig. 3. It can be seen that the average signal amplitude that came to the PMT from the far end of the spectrometer (insert in Fig. 3) is about half the average amplitude for all values of the voltage in the case of passing the particle near the PMT. Inhomogeneity in the amplitudes' values along the spectrometer length exists for almost all the studied values of voltage on the PMT divider. However, at a voltage U = 1100 V at ~90% of the spectrometer length, the amplitude value is almost constant. The signals inhomogeneity at a significant spectrometer depth shows that in extended and rather "narrow" volume of the spectrometer, the light collection is affected by a geometric factor. The value of the geometric factor can be influenced by changing the size of the spectrometer. In this case, it cannot be done, but its effect can be reduced by desensitization of the PMT, i.e. by reducing the voltage on the divider (U ≤ 1180 V) (dependence 1 in Fig. 3).

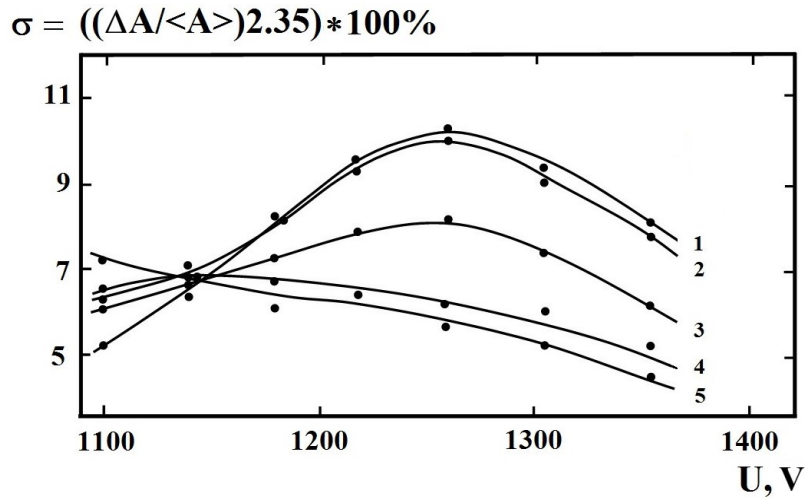

**Fig. 4.** Dependence of the energy resolution of the extended quadrahedral NaI(Tl) spectrometer`s from the voltage on the PMT-52 voltage divider and the entry point of the cosmic muon into the spectrometer (insert Fig. 3): x = 4 cm; x = 22 cm; x = 38 cm; x = 40 cm; x = 44 cm.

The dependence of the NaI(Tl) spectrometer's energy resolution on its length and voltage at the PMT divider is shown in Fig. 4 ($\sigma = ((\Delta A/<A>)/2.35)\cdot 100\%$, where $\Delta A$ is the full width of the amplitude spectrum at half its height; $<A>$ is the average amplitude of the spectrum). It can be seen that σ varies in the range from 4% to 10% depending on the voltage at the divider and the point of muon passage through the spectrometer. The resolution deteriorates with distance of the muon passage point through the spectrometer from the PMT and the voltage increase on the divider. This is probably because, in the volume regions of the spectrometer far from the PMT, there are significant fluctuations associated with the light collection in the dense substance of the spectrometer. However, Fig. 4 shows that there is a small voltage range U = 1150±10 V, at which the energy resolution of the NaI(Tl) spectrometer is almost constant over the entire length of the spectrometer and is σ ≈ 7%. It can be assumed that for these spectrometer sizes and the type of used PMT, the influence of the geometric factor on the energy characteristics of the spectrometer at a voltage U = 1150±10 V is minimal.

The dependencies were obtained similar to those shown in Fig. 3 and 4during the calibration on cosmic muons of the three remaining NaI(Tl) spectrometers. The numerical values of the obtained dependencies differ from the presented ones by no more than ~20%.